\documentclass[doublespacing]{elsart}

\usepackage{amssymb}
\usepackage{amsmath}
\usepackage{bm}
\usepackage{graphicx}
\begin{document}

\begin{frontmatter}



\title{Quadrupole Moments of Neutron-Deficient $^{20, 21}$Na}


\author[triumf,nscl]{K. Minamisono\thanksref{label1}}\ead{minamiso@nscl.msu.edu},
\author[osaka]{K. Matsuta},
\author[osaka]{T. Minamisono\thanksref{label2}},
\author[triumf]{C. D. P. Levy},
\author[osaka]{T. Nagatomo\thanksref{label3}},
\author[osaka]{M. Ogura},
\author[osaka]{T. Sumikama\thanksref{label4}},
\author[triumf]{J. A. Behr},
\author[triumf]{K. P. Jackson},
\author[osaka]{M. Mihara}, and
\author[osaka]{M. Fukuda},

\address[triumf]{TRIUMF, 4004 Wesbrook Mall, Vancouver, BC V6T 2A3, Canada}
\address[nscl]{National Superconducting Cyclotron Laboratory, Michigan State University, East Lansing, MI 48824, USA}
\address[osaka]{Department of Physics, Osaka University, Toyonaka, Osaka 560-0043, Japan}

\thanks[label1]{Japan Society for the Promotion of Science, Postdoctoral Fellowships for Research Abroad}
\thanks[label2]{Current address: Department for the Application of Nuclear Technology, Fukui University of Technology, Gakuen, Fukui 910-8505, Japan}
\thanks[label3]{Current address: RIKEN, 2-1 Hirosawa, Wako, Saitama 351-0198, Japan}
\thanks[label4]{Current address: Department of Physics, Tokyo University of Science, Noda, Chiba 278-8510, Japan}

\begin{abstract}
The electric-quadrupole coupling constant of the ground states of the proton drip line nucleus $^{20}$Na($I^{\pi}$ = 2$^{+}$, $T_{1/2}$ = 447.9 ms) and the neutron-deficient nucleus $^{21}$Na($I^{\pi}$ = 3/2$^{+}$, $T_{1/2}$ = 22.49 s) in a hexagonal ZnO single crystal were precisely measured to be $|eqQ/h| = 690 \pm 12$ kHz and 939 $\pm$ 14 kHz, respectively, using the multi-frequency $\beta$-ray detecting nuclear magnetic resonance technique under presence of an electric-quadrupole interaction.  A electric-quadrupole coupling constant of $^{27}$Na in the ZnO crystal was also measured to be $|eqQ/h| = 48.4 \pm 3.8$ kHz.  The electric-quadrupole moments were extracted as $|Q(^{20}$Na)$|$ = 10.3 $\pm$ 0.8 $e$ fm$^2$ and $|Q(^{21}$Na)$|$ = 14.0 $\pm$ 1.1 $e$ fm$^2$, using the electric-coupling constant of $^{27}$Na and the known quadrupole moment of this nucleus as references.  The present results are well explained by shell-model calculations in the full $sd$-shell model space.  
\end{abstract}

\end{frontmatter}

% main text
\section{Introduction}
\label{section:1}
The spectroscopic electric-quadrupole moment ($Q$) provides a direct measure of the deviation of the charge distribution in a nucleus from spherical symmetry and thus is sensitive to details of the nuclear wavefunction.  $Q$ is often used in tests of various theories, which attempt to reproduce experimental data.  Shell-model calculations have been successful in predicting $Q$ near the stability line in the nuclear chart, where the required effective interactions are well constrained by experimental data (for example, Ref. \cite{brown06}).  Proton and neutron effective charges are usually required in shell-model calculations to obtain $Q$.  Within a given shell-model space, the effective charges represent 2$\hbar \omega$, $I^{\pi}$ = 2$^+$ excitations of the core nucleons to valence orbits \cite{sagawa84} and reflect the virtual excitation of the isoscalar and isovector giant quadrupole resonances.  Values of effective charges for $sd$-shell nuclei have been obtained from a systematic analysis of experimental $E$2 matrix elements \cite{sagawa84, brown88}. 

When moving away from stability, however, some nuclei show significant disagreement between experiment and theory.  The neutron-rich B isotopes, for example, show a reduction of the neutron effective charge \cite{ogawa04} because the loosely-bound valence neutrons, far removed from the core, have less probability to excite the core than well bound neutrons.  Such variation of the neutron effective charge can also be seen in the neutron-rich $^{16}$N \cite{matsuta01} and $^{18}$N \cite{ogawa99}.  Neutron-deficient nuclei, especially those with small proton separation energies, may also be expected to show variation of effective charges.  One such example is the $Q$ of neutron-deficient $^{37}$K \cite{minamisono08b}, which requires an increased neutron effective charge and a decreased proton effective charge relative to the typical values of nuclei closer to stability in the $sd$ shell.  This is attributed to a substantial coupling to the isovector giant resonance beyond that indicated by the typical effective charges.  Experimental $Q$ of such neutron-deficient nuclei are still scarce, even in the $sd$ shell.  Additional systematic data of $Q$ of neutron-deficient nuclei are important to further improve our knowledge of the exotic structure of drip line nuclei.  

The $Q$ of the neutron-rich Na isotopes have been measured via laser spectroscopy \cite{touchard82} and in $\beta$-ray detecting nuclear magnetic resonance ($\beta$ NMR) \cite{keim00} experiments.  Intrusion of the $fp$ shell across the $N$ = 20 shell gap can be clearly seen in $Q$ of ground states of neutron-rich Na isotopes beyond $^{29}$Na in the deviation between experiment and shell model calculations restricted to the $sd$-shell model space \cite{utsuno04}.  Improved agreement is realized in the Monte Carlo Shell Model approach, where both the $sd$- and $fp$-shell model spaces are considered \cite{utsuno04}.  Compared with the well studied neutron-rich Na isotopes, precise $Q$ for the neutron-deficient Na isotopes are still lacking.  The $Q$ of the proton drip line nucleus $^{20}$Na($I^{\pi}$ = 2$^{+}$, $T_{1/2}$ = 447.9 ms) was reported in a figure in Ref. \cite{keim00b} but no experimental detail nor a value of $Q$ was given.  The $Q$ of neutron-deficient nucleus $^{21}$Na($I^{\pi}$ = 3/2$^{+}$, $T_{1/2}$ = 22.49 s) disagrees with a shell-model calculation in the $sd$-shell model space [$Q_{\rm theor.}(^{21}$Na) = 11.6 $e$ fm$^2$ (calculated using OXBASH shell model code \cite{oxbash}], although it has a large error [$Q(^{21}$Na) = 6 $\pm$ 3 $e$ fm$^2$ (reevaluated value using Ref. \cite{touchard82,wo93,jonsson96})].

$Q(^{20}$Na) and $Q(^{21}$Na) have been precisely determined in this study using the multi-frequency $\beta$-NMR technique under presence of an electric-quadrupole interaction.  Preliminary results were reported in Ref. \cite{ogura04}.  %The present $|Q(^{20}$Na)$|$ = 10.3 $\pm$ 0.8 $e$ fm$^2$ reproduces and confirms the previous report \cite{keim00b}.  The present $|Q(^{21}$Na)$|$ = 14.0 $\pm$ 1.1 $e$ fm$^2$ reconciles the disagreement between experimental and theoretical $Q(^{21}$Na) arising from the previous result.  
The systematic behavior of $Q$(Na) across the full neutron $sd$ shell, including the present results of $Q(^{20}$Na) and $Q(^{21}$Na), is discussed.

\section{EXPERIMENT}
The experiments were performed at the radioactive beam facility ISAC-I at TRIUMF.  The electric-quadrupole coupling constants, $eqQ/h$, of $^{20, 21, 27}$Na were measured in separate runs.  The experimental procedure in the measurement of $^{20}$Na is explained below.  Similar procedures were taken in the measurements of $^{21, 27}$Na and the minor changes in conditions among the three measurements are summarized in Table \ref{table:expcondition}.  The 500 MeV proton beam from the TRIUMF cyclotron was used to bombard a thick SiC production target, which was coupled to a surface ion source.  $^{20}$Na singly-charged ions were extracted at an energy of 40.8 keV and mass separated.  The pure $^{20}$Na beam was transported to the polarizer beam line \cite{levy02} in the ISAC-I experimental hall.  The $^{20}$Na ions first passed through a Na vapor and were neutralized by charge exchange reactions.  Collinear laser pumping was used to polarize the Na atoms by the D$_1$ transition ($3s$ $2S_{1/2} \leftrightarrow 3p$ $2P_{1/2}$) and circularly polarized light \cite{arnold87}.  Both of the ground state hyperfine levels ($3s$ $2S_{1/2}$ $F = I + 1/2$ and $I - 1/2$) were pumped to achieve high polarization (see Table \ref{table:expcondition}) using side band frequencies produced by an electro-optic modulator (EOM), a technique that was successfully employed in the past \cite{levy03}.  The collinear laser light was generated by a Coherent 899-21 frequency-stabilized ring-dye-laser pumped by a 7-W argon-ion laser.  The polarized atoms were then re-ionized in a He-gas target to be deflected to the $\beta$-NMR station.

The polarized $^{20}$Na ions were delivered to the $\beta$-NMR apparatus and implanted into a hexagonal ZnO single crystal.  The implantation depth was $\sim$ 500\AA.  An external dipole-magnetic field of $B_0$ = 0.5286 $\pm$ 0.0005 T was applied parallel to the direction of polarization to maintain the polarization in ZnO and to make the magnetic sublevels split (Zeeman splitting).  The $^{20}$Na nucleus decays mainly to the first excited state in daughter nucleus $^{20}$Ne by emitting $\beta^+$ rays with a half-life of 447.9 ms.  The maximum $\beta$-ray energy is 11.23 MeV.  $\beta$ rays from the stopped $^{20}$Na were detected by a set of plastic scintillation counters placed at 0$^\circ$(u) and 180$^\circ$(d) relative to the external field direction.  The counting rate is asymmetric between u and d counters for a polarized source.  The angular distribution,
\begin{equation}
\label{eq:angulardistribution}
W(\varphi) \sim 1+A_sP\cos{\varphi},
\end{equation} 
depends on the asymmetry parameter $A_s$, the polarization $P$ and the angle $\varphi$ between the direction of the momentum of the $\beta$ ray emitted in the decay and the polarization axis.  The $A_s$ for the $\beta$ decay of $^{20}$Na to the first excited state in $^{20}$Ne is $A_s$ = 1/3.  The initial magnitude ($P_0$) and spin-lattice relaxation time ($T_1$) of the nuclear polarization of $^{20}$Na were measured in this study from the asymmetric $\beta$-ray angular distribution.  $P_0$ = 37 $\pm$ 1\% and $T_1$ = 9 $\pm$ 0.5 s were obtained for $^{20}$Na in ZnO.  The preliminary results of production of polarization were reported in Ref. \cite{minamisono04}.  The long $T_1$ relative to the half-life introduced virtually no significant loss of polarization in the NMR measurement.  

The ZnO was chosen for implantation to measure $eqQ/h$ because polarization of Na isotope is well maintained in the ZnO with long relaxation time \cite{minamisono04} relative to their lifetimes.   The ZnO had its c-axis set perpendicular to the external magnetic field.  The electric-field gradient, $q$, in ZnO is axially symmetric (asymmetry parameter $\eta = 0$) and parallel to the c-axis.  The Hamiltonian of the electromagnetic interaction between nuclear moments and external fields \cite{abragam86} in this condition is given by
\begin{equation}
H = -\bm{\mu}\cdot\bm{H}_0+\frac{eqQ}{4I(2I-1)}\{3I_Z^2-I(I+1)\},
\label{eq:hamiltonian}
\end{equation}
where $\bm\mu$ is the magnetic moment, $\bm{H}_0$ is  the magnetic field, $I$ is the nuclear spin and $I_z$ is the third component of the spin operator.  %The Hamiltonian with $\eta$ = 0 can be given in a simpler form than that with finite value of $\eta$, which reduces a possible systematic error in determining $eqQ/h$ resulting from  calculated resonance frequencies with experimental error in $\eta$.  
The magnetic sublevel energies, $E_m$, of the Na ions implanted in ZnO are given by 
\begin{equation}
E_m = -g\mu_NH_0m + \frac{h\nu_Q}{12}(3\cos^2\theta-1)\{3m^2-I(I+1)\},
\label{eq:energy}
\end{equation}
where $m$ is the magnetic quantum number, $g$ is the nuclear $g$ factor, $\nu_Q = 3eqQ/\{2I(2I-1)h\}$ is the normalized electric-quadrupole frequency and $\theta$ is the angle between the c-axis and the external magnetic field.  Eq. (\ref{eq:energy}) is given to first-order in $eqQ/h$, taking the electric-quadrupole interaction as a perturbation to the main magnetic interaction.  The first term in Eq. (\ref{eq:energy}) involves the $2I + 1$ magnetic sublevels separated by a fixed energy value due solely to the magnetic interaction (Zeeman splitting).  These sublevels are further shifted by the electric-quadrupole interaction and the energy spacing between adjacent sublevels is no longer constant.  The 2$I$ separate transition frequencies that appear due to the electric-quadrupole interaction are determined as
\begin{equation}
\nu_{m-1\leftrightarrow m} = \nu_L - \frac{\nu_Q}{4}(3\cos^2\theta-1)(2m-1).
\label{eq:frequency}
\end{equation}
The transition frequencies correspond to the energy difference between two adjacent levels in Eq. (\ref{eq:energy}), since the allowed transitions have $\Delta m = \pm1$.  Here $\nu_L = g\mu_NH_0/h$ is the Larmor frequency.  Precise determination of $\nu_L$ is important for an accurate measurement of $eqQ/h$, since the 2$I$ transition frequencies appear relative to the location of $\nu_L$ as seen in Eq. (\ref{eq:frequency}).  The $\nu_L$ was measured in a cubic NaF single crystal in the present study, prior to the $eqQ/h$ measurements.  The use of the perturbation technique to obtain transition frequencies is justified for $^{21}$Na in ZnO and $^{27}$Na in ZnO because the perturbation condition, $\nu_L \gg \nu_Q$, is satisfied.  On the other hand, for $^{20}$Na in ZnO, the Hamiltonian [Eq. (\ref{eq:hamiltonian})] was numerically solved to obtain transition frequencies because the perturbation technique cannot be applied, due to the condition $\nu_L \sim \nu_Q$.  %However, the principle of magnetic-sublevel splittings is essentially the same as discussed above, when the electric quadrupole interaction is superimposed on the magnetic interaction.  

An asymmetry change, $A_sP$, in the $\beta$-decay angular distribution is obtained as the NMR signal: \begin{equation}
A_sP = \frac{\sqrt{R}-1}{\sqrt{R}+1}.
\label{eq:ap}
\end{equation}  
The double ratio, $R$, is defined by $\beta$-ray countings, $W(\theta)$, as 
\begin{equation}
R = \left[\frac{W(0^\circ)}{W(180^\circ)}\right]_{\rm off}\Bigg/\left[\frac{W(0^\circ)}{W(180^\circ)}\right]_{\rm on},
\label{eq:udratio}
\end{equation} 
where the subscript, off(on), stands for without(with) radio frequency (RF) applied.  An adiabatic fast passage (AFP) method \cite{abragam86b} was used in the RF application for NMR.  AFP inverts the direction of initial polarization ($P \rightarrow -P$) and effectively doubles the signal size over a depolarization method for a more efficient NMR measurement.  

The NMR signal can be maximized and therefore effectively searched by applying all the 2$I$ transition frequencies in Eq. (\ref{eq:frequency}) in a short time period compared with the lifetime of the nucleus.  Such application ensures a saturation of all transitions in Eq. (\ref{eq:frequency}) at $eqQ/h$.  The NMR signal obtained for a nucleus with $I$ = 2 (as is the case for $^{20}$Na) in the AFP method with multiple transition frequencies is 20 times bigger than that obtained by applying a single transition frequency, where only a partial saturation of transitions can be achieved.  In this comparison, a linear distribution of populations in $E_m$ is assumed as $a_m - a_{m-1} = \epsilon$ and $\sum^{I}_{m=-I}a_m = 1$, where $a_m$ is the population in $m$ and $\epsilon$ is a constant.  This multi-frequency $\beta$-NMR technique is discussed in detail elsewhere \cite{minamisono93, minamisono08}.

\begin{figure}[h]
\begin{center}
\includegraphics[width=12cm,keepaspectratio,clip]{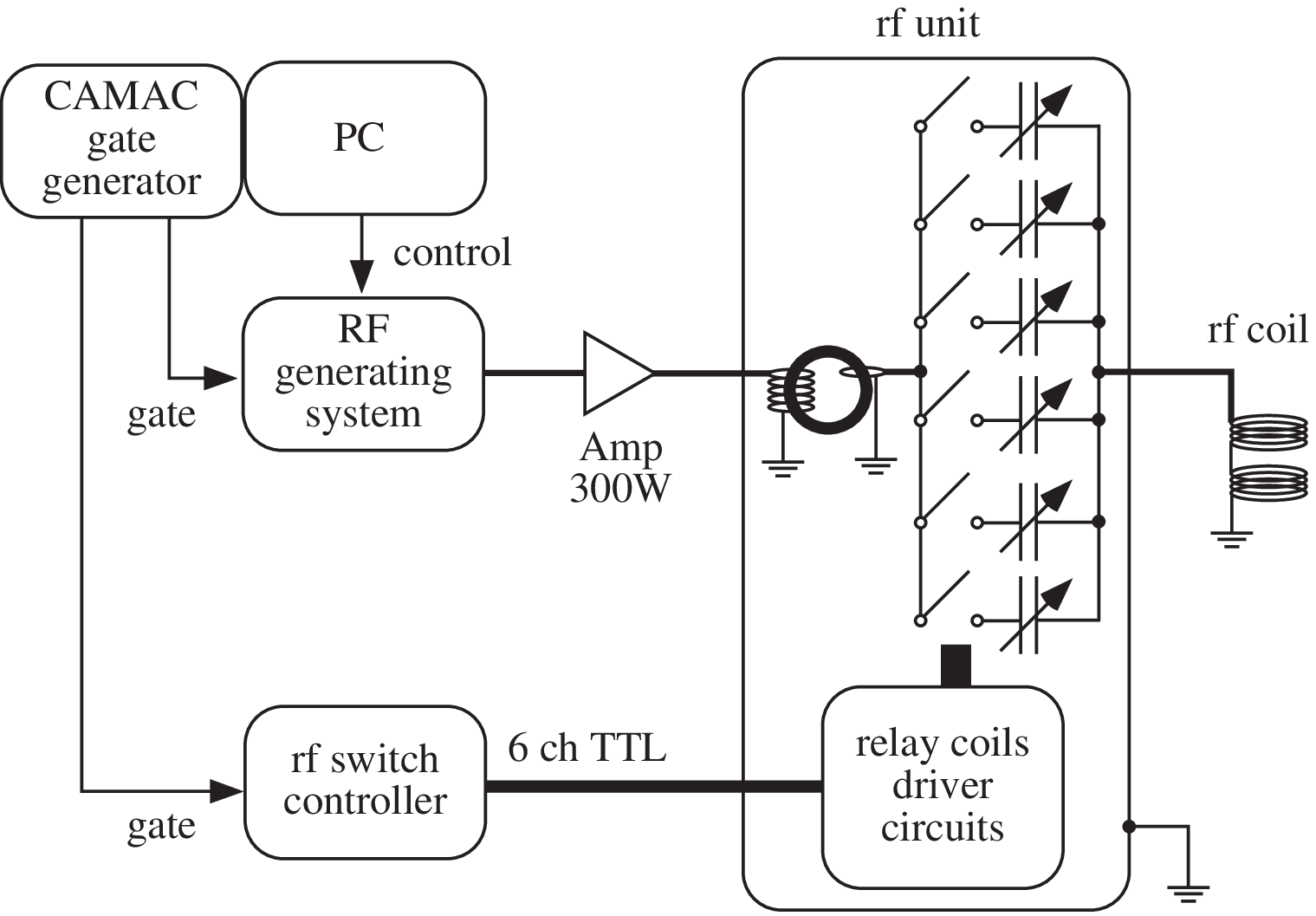}
\end{center}
\caption{Schematic illustration of the multi-frequency $\beta$-NMR system.  The rf coil was placed in a vacuum chamber, which is not drawn here.}
\label{fig:rfcircuit}
\end{figure}

A schematic of the multi-frequency $\beta$-NMR system is shown in Fig. \ref{fig:rfcircuit}.  A computer controlled RF generating system produces an RF signal, which is sent to a 300 W amplifier.  The amplified signal is applied to an RF coil, which is part of an LC resonance circuit.  The circuit includes an impedance matching transformer and a bank of six selectable variable capacitors.  After 10 ms RF time for $^{20}$Na, another frequency is generated by the RF generating system and sent to the same LC resonance circuit.  A different capacitor, which has been tuned to the second frequency to satisfy the LC resonance condition, is selected by the fast switching relay system.  The system ensures sufficient power for any set of four transition frequencies for $^{20}$Na ($I$ = 2) over the expected search region of $eqQ/h$.  The switching time between two signals is 3 ms.

\begin{figure}[h]
\begin{center}
\includegraphics[width=12cm,keepaspectratio,clip]{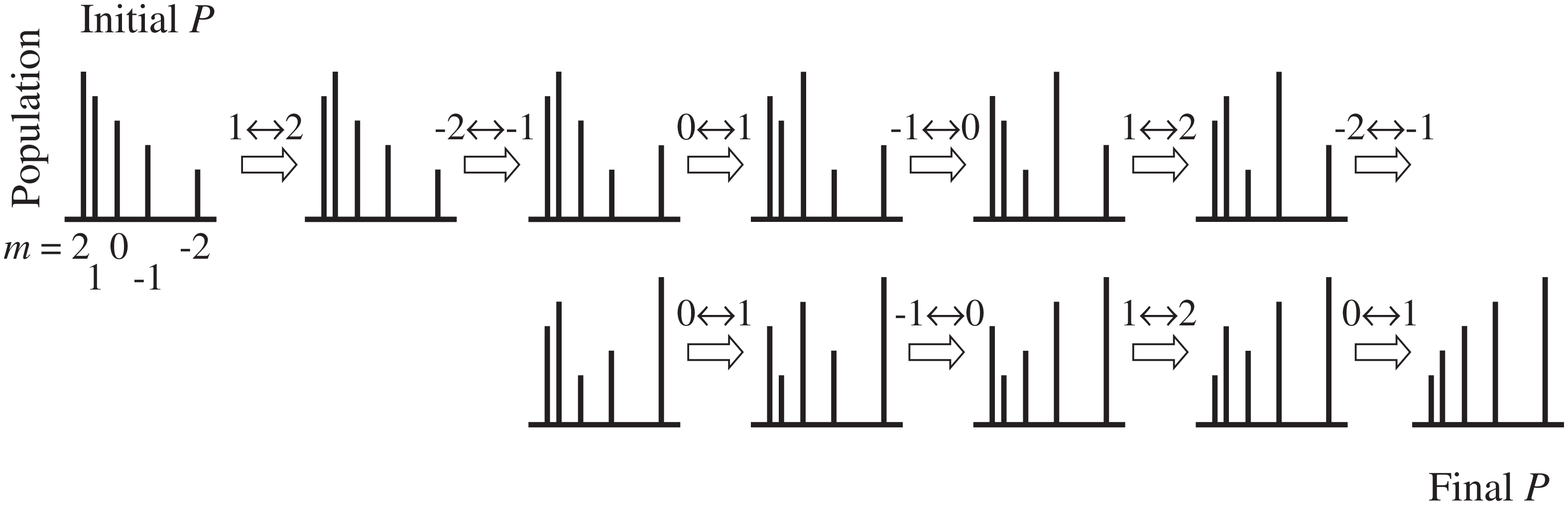}
\end{center}
\caption{Schematic illustration of the inversion of initial polarization by AFP for an $I$ = 2 nucleus.  1 $\leftrightarrow$ 2, for example, indicates an application of transition frequency between $m$ = 1 and 2 and the populations in these two sublevels are interchanged by the AFP.  Note that the spacings among energy levels are uneven due to the electric-quadrupole interaction.}
\label{fig:spinmanipulation}
\end{figure}

%The four transition frequencies of $^{20}$Na in ZnO for certain $eqQ/h$ calculated using Eq. (\ref{eq:frequency}) were sequentially applied through the RF coil.  
Since an RF corresponds to a transition of populations between adjacent $m-1$ and $m$, the AFP interchanges the $a_{m-1}$ and $a_{m}$.  The total inversion of polarization ($P \rightarrow -P$)  was achieved by applying RF as shown in Fig. \ref{fig:spinmanipulation} for $^{20}$Na ($I$ = 2).  %In the figure, $1 \leftrightarrow 2$ corresponds to an application of the transition frequency between $m$ = 1 and 2 and the populations in these two sublevels are interchanged by the AFP.  
After 10 sequential applications of transition frequencies, the direction of initial polarization is inverted, where $P$ is defined as $P = \sum^{I}_{m=-I} a_mm/I$.  The typical inversion efficiency, defined by $P^\prime = \alpha P_0$, was $\alpha \sim -0.81$, where $P^\prime$ is the inverted polarization.   Each applied RF was frequency modulated with a modulation width, FM = $\pm$ 20 kHz, for AFP.  The FM is also to cover a certain region of $eqQ/h$ for an effective search for resonance.  The FM was swept only once in the RF time of 10 ms.  An amplitude modulation was also applied to the RF for an efficient AFP.  The oscillating magnetic field strength, $H_1$, was $\sim$ 0.8 mT at the maximum.  At the beginning (end) of the FM sweep, the $H_1$ was reduced to $\sim$ 0.02 mT and gradually increased (decreased) to the maximum (minimum) $H_1$ to avoid depolarization due to sudden start (stop) of the strong $H_1$.  A pulsed beam method was employed using a fast electrostatic kicker.  A 500 ms implantation time was followed by the RF time and then by a 1 s $\beta$-ray counting time.  The sequence was repeated until a required statistics was achieved.  A 10 s implantation time and 35 s counting time were used in the $^{21}$Na measurement.  Only the first 5 s counting time was used to calculate the NMR signal of $^{21}$Na, because of the short relaxation time in ZnO compared with the lifetime of $^{21}$Na as noted in Table \ref{table:expcondition}.

Multi-frequency $\beta$-NMR spectra were measured both with a positive-helicity ($\sigma^+$) and a negative-helicity ($\sigma^-$) laser light.  %A small disagreement between applied and actual-resonance frequencies was able to be detected, if any, as asymmetric amplitude and/or shape of resonances between spectra measured with $\sigma^+$ and $\sigma^-$.  By adjusting the applied frequencies for the spectra to be symmetric, accurate measurements were possible.  
An NMR signal, 2$A_sP_{av}$, was defined as:
\begin{equation}
2A_sP_{av} \equiv A_sP(\sigma^+) - A_sP(\sigma^-),
\label{eq:effectiveap}
\end{equation}
to maximizes the NMR signal obtained.

\section{RESULTS}
\begin{figure}[h]
\begin{center}
\includegraphics[width=12cm,keepaspectratio,clip]{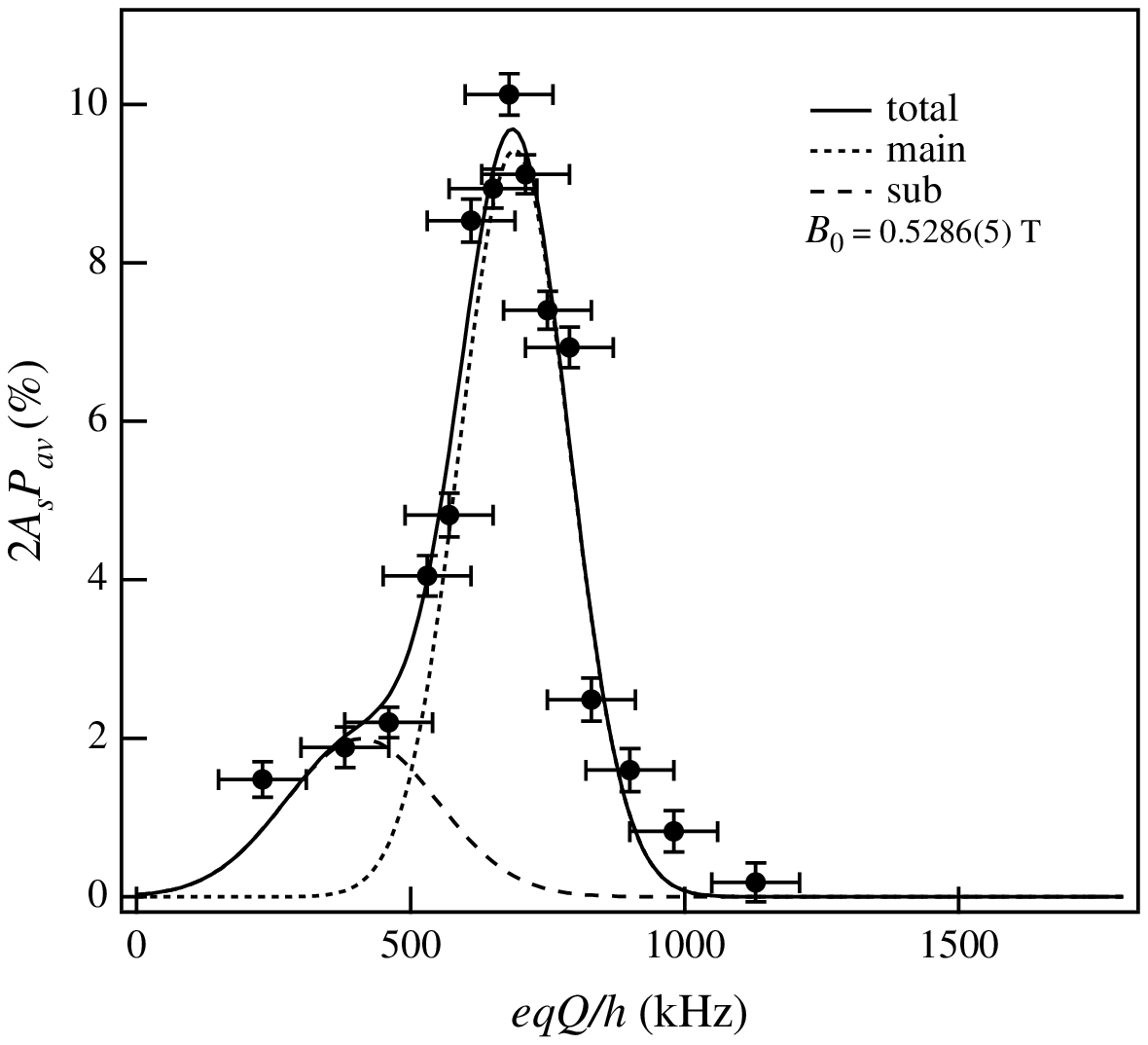}
\end{center}
\caption{The multi-frequency $\beta$-NMR spectrum of $^{20}$Na in ZnO.  The solid line is the two Gaussian fit.  The dotted and the dashed lines are each component of the total fit.  The sub resonance was fixed by the fitting result of $^{21}$Na in ZnO.}
\label{fig:20NainZnO}
\end{figure}

\begin{figure}[h]
\begin{center}
\includegraphics[width=12cm,keepaspectratio,clip]{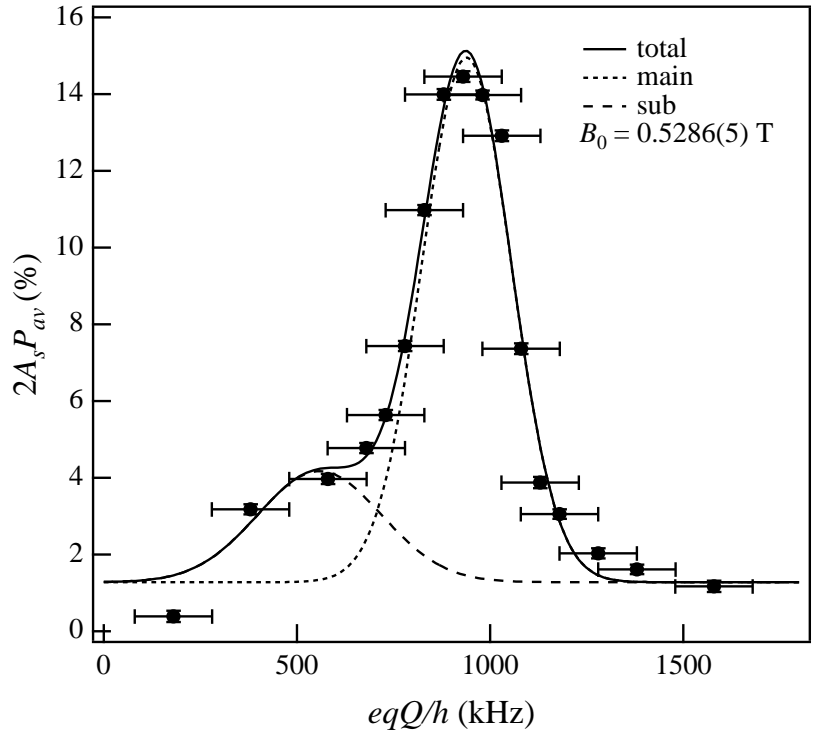}
\end{center}
\caption{The multi-frequency $\beta$-NMR spectrum of $^{21}$Na in ZnO.  The solid line is the two Gaussian fit.  The dotted and the dashed lines are each component of the total fit.}
\label{fig:21NainZnO}
\end{figure}

The multi-frequency $\beta$-NMR spectra of $^{20}$Na and $^{21}$Na in ZnO are shown in Fig. \ref{fig:20NainZnO} and \ref{fig:21NainZnO}, respectively.  The solid circles are the experimental data and the horizontal bar on each point is the range of $eqQ/h$ covered by the FM.  There appears to be two final locations of Na ions in ZnO, where the crystal lattice provides a different electric-field gradient on Na.  The resonance peak at higher $eqQ/h$ is named ``main'' (the dotted line) and one at lower $eqQ/h$ ``sub'' (the dashed line).  The measurement of the sub resonance of $^{20}$Na in ZnO was incomplete as seen in Fig. \ref{fig:20NainZnO}, as no data points were taken below $eqQ/h \sim 200$ kHz.  Therefore, the ratios of the centroids (namely the ratio of electric-field gradients), the amplitudes ($k$) and the line widths ($\sigma$) between the main and sub resonances were taken from the fit of $^{21}$Na in ZnO.  The central values of $q_s/q_m$ = 0.59 $\pm$ 0.01, $k_s/k_m$ = 0.21 $\pm$ 0.01 and $\sigma_s/\sigma_m$ = 1.4 $\pm$ 0.1, where the subscripts $s$ and $m$ refer to sub and main, respectively, were used as fixed parameters in fitting of the $^{20}$Na resonance structure.  A baseline shift of (1.3 $\pm$ 0.1)\% was observed in 2$A_sP_{av}$ in the spectrum of $^{21}$Na in ZnO [see Fig. \ref{fig:21NainZnO}].  This is because the transition between $m$ = $-$1/2  $\leftrightarrow$1/2 coincides with the Larmor frequency ($I$ = 3/2) within the searched region of $eqQ/h$, where the first perturbation calculation is valid.  As a result, the transition is always induced by RF, and 1/10 of the maximum 2$A_sP_{av}$ appears as a shift of baseline for $I = 3/2$ nucleus, assuming a distribution of magnetic sublevel populations.  The fit results give $eqQ/h$ of the main resonances of $^{20}$Na and $^{21}$Na in ZnO:
\begin{align}
\left|\frac{eqQ(^{20}{\rm Na})}{h}\right| &= 690 \pm 2({\rm stat.}) \pm 12({\rm syst.}) {\rm \ \ kHz}
\label{eq:couplingconst20},\\
\left|\frac{eqQ(^{21}{\rm Na})}{h}\right| &= 939 \pm 2({\rm stat.}) \pm 14({\rm syst.}) {\rm \ \ kHz}.
\label{eq:couplingconst21}
\end{align}
The second error is the systematic error in determining the centroid.  Possible variations of the centroid due to the sub resonance were considered in the systematic error based on the errors of $q_s/q_m$, $k_s/k_m$ and $\sigma_s/\sigma_m$.  One tenth of the $\sigma_m$ was also included in the systematic error \cite{minamisono98} for uncertainties caused by the line width of main resonance due to the FM.  The latter dominates the present systematic error.

\begin{figure}[h]
\begin{center}
\includegraphics[width=12cm,keepaspectratio,clip]{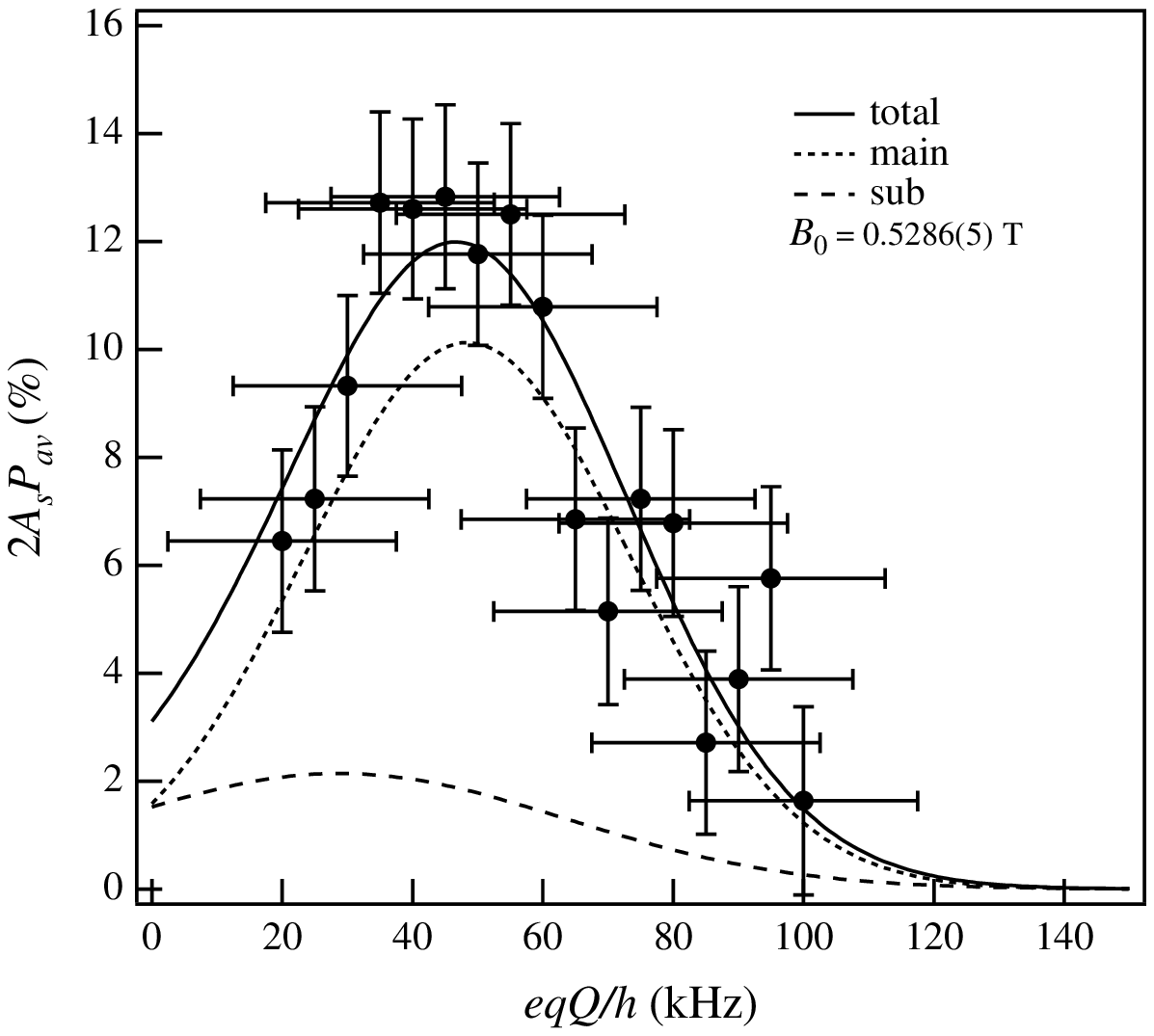}
\end{center}
\caption{The multi-frequency $\beta$-NMR spectrum of $^{27}$Na in ZnO.  The solid line is the total fit.  The dotted and the dashed lines are each component of the total fit.  The sub resonance was fixed by the fitting result of $^{21}$Na in ZnO.}
\label{fig:27NainZnO}
\end{figure}

$Q$ may be extracted from these $eqQ/h$ together with an $eqQ/h$ of a Na isotope in ZnO, whose $Q$ is already known as the electric-field gradient is identical among Na isotopes.  The $eqQ/h$ of $^{27}$Na in ZnO was measured for this purpose.  The $Q$ is precisely known in a separate measurement as $Q(^{27}$Na) = 0.72 $\pm$ 0.03 $e$ fm$^2$ \cite{keim00}.  The multi-frequency $\beta$-NMR spectrum of $^{27}$Na in ZnO is shown in Fig. \ref{fig:27NainZnO}.  Since the $eqQ/h$ is small, the main and sub resonances were not resolved.  The fitting of a two-component Gaussian was performed, following the same procedure as outlined previously for the analysis of $^{20}$Na in ZnO.  The transition between $m$ = $-$1/2  $\leftrightarrow$1/2 was always induced ($I$ = 5/2) by RF.  The expected NMR effect by the transition is 3/100 of the maximum 2$A_sP_{av}$, therefore the baseline of the Gaussians was fixed to 2$A_sP_{av}$ = 0\% in the fitting, which introduced no significant systematic error.  The centroid of the fit gives the $eqQ/h$ of main resonance of $^{27}$Na in ZnO:
\begin{equation}
\left|\frac{eqQ(^{27}{\rm Na})}{h}\right| = 48 \pm 2({\rm stat.}) \pm 3({\rm syst.}) {\rm \ \ kHz}.
\label{eq:couplingconst27}
\end{equation}
Possible variations of the centroid, due to the sub resonance as well as a one tenth of the $\sigma_m$ were included in the systematic error.  The $Q$ of Na isotope with a mass number $A$ can now be extracted as
\begin{equation}
Q(^{A}{\rm Na}) = \left[\frac{eqQ(^{A}{\rm Na})}{h}\Bigg/\frac{eqQ(^{27}{\rm Na})}{h}\right]_{\rm ZnO} \times Q(^{27}{\rm Na}).
\label{eq:eqqratio}
\end{equation}   
%The $eqQ/h$ of main resonances will be discussed in the paper to extract $Q$ since the sub resonances in the $^{20}$Na and $^{27}$Na in ZnO spectra were not resolved.

$Q(^{20}$Na) and $Q(^{21}$Na) were precisely extracted from Eq. (\ref{eq:eqqratio}) and $Q(^{27}$Na) \cite{keim00} as
\begin{align}
|Q(^{20}{\rm Na})| &= 10.3 \pm 0.8\ \ e{\rm \ \ fm}^2,\\
|Q(^{21}{\rm Na})| &= 14.0 \pm 1.1\ \ e{\rm \ \ fm}^2.
\end{align}
The statistical and systematic errors were added quadratically.  The results are summarized in Table \ref{table:expresult} together with the previously known $Q$ and the shell-model predictions discussed in the following section.

\section{DISCUSSION}
\begin{figure}[h]
\begin{center}
\includegraphics[width=12cm,keepaspectratio,clip]{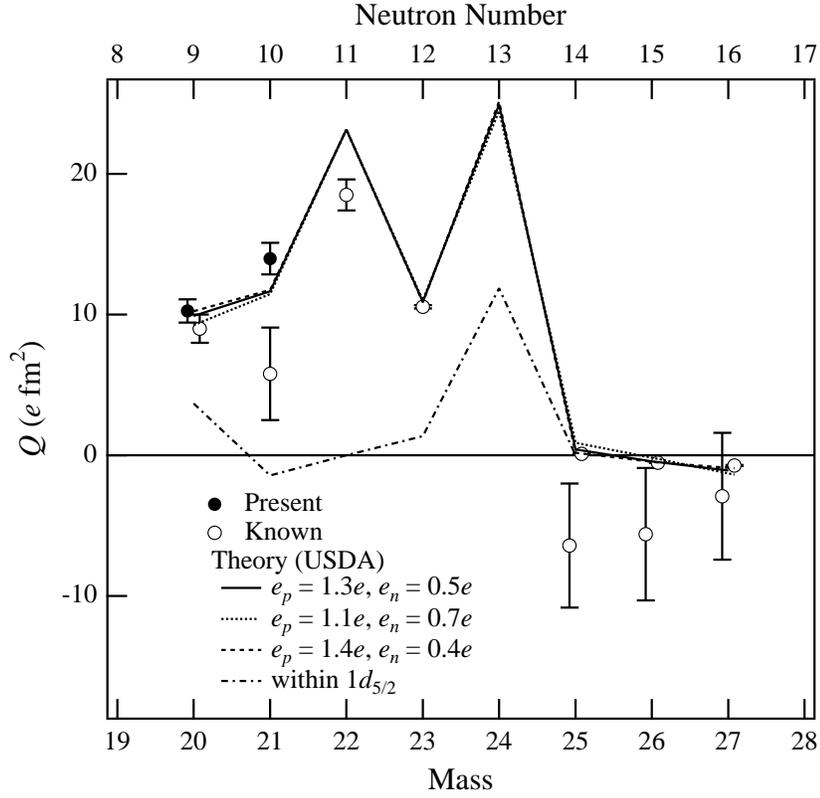}
\end{center}
\caption{$Q$ of the Na isotopes below $A$ = 28.  The solid circles are the present data, for which signs are taken from the shell-model calculations. The open circles are known experimental data.  See Table \ref{table:theory} for the values.  The solid line shows the shell model results with the USDA interaction \cite{brown06}.}
\label{fig:naqmom}
\end{figure}

The present $Q(^{20}$Na) is consistent with and as precise as the previous report \cite{keim00b} measured by the $\beta$-NMR technique.  The experimental $Q(^{20}$Na) is well confirmed.  The present $Q(^{21}$Na) is not consistent though more precise than the previous value, $Q(^{21}$Na) = 6 $\pm$ 3 $e$ fm$^2$.  The previous $Q(^{21}$Na) was obtained from the hyperfine coupling constant of the 3$p$ $^2P_{3/2}$ state in $^{21}$Na ($B_{P_{3/2}}$ = 1.5 $\pm$ 0.8 MHz).  The value of $B_{P_{3/2}}$ was a reevaluated value using the hyperfine splitting of 3$p$ $^2P_{3/2}$ state in $^{21}$Na ($\nu_0(F=3 \leftrightarrow 2) = 61.4 \pm 1.0$ MHz \cite{touchard82}), the latest values of the hyperfine coupling constant of $^{23}$Na ($A_{P_{3/2}}$ =  18.534 $\pm$ 0.015 MHz \cite{wo93}), and $Q$ of $^{23}$Na ($Q(^{23}$Na) = 10.6 $\pm$ 0.1 $e$ fm$^2$ \cite{jonsson96}).  The large error in the previous $Q(^{21}$Na) is due to the large error in $\nu_0(F=3 \leftrightarrow 2)$, which is essentially a systematic error caused by the detuning of the laser frequency from the exact resonance \cite{touchard82}.  The deviation between present and previous $Q(^{21}$Na) may reside in the systematic error.  It is noted that the similar discrepancy between $Q$ measured by $\beta$-NMR \cite{keim00} and optical \cite{touchard82} technique is systematically seen in other Na isotopes, which was shown in Fig. 5 of Ref. \cite{keim00} and also in Fig. \ref{fig:naqmom}.

Theoretical calculations were performed using OXBASH shell model code \cite{oxbash} in the full $sd$-shell model space from $A$ = 20 to 27 with the USDA interaction \cite{brown06} and Woods-Saxson single-particle wave functions.  Calculations above $A$ = 28 were not included, although $Q$ are known up to $A$ = 31 \cite{keim00b, wilbert00}.  The contribution from 2p-2h intruder configuration across the $N$ = 20 shell gap becomes important above $^{29}$Na, and these nuclei therefore do not compare well with shell-model calculations in the $sd$-shell model space \cite{utsuno04}.  Theoretical $Q$ were calculated with $Q_{\rm theory} = e_pQ_0^p + e_nQ_0^n$, where $Q_0^{p(n)}$ is the bare $Q$ of the proton (neutron) distribution and $e_p$ and $e_n$ are the effective charges for protons and neutrons, respectively.  Values of $e_p$ = 1.3$e$ and $e_n$ = 0.5$e$ were used for the shell-model results.  These effective charges are typical for $sd$ shell nuclei \cite{sagawa84, brown88}.  The results of theoretical calculations and experiments are summarized in Table \ref{table:theory} and shown in Fig. \ref{fig:naqmom}, where signs of present results (the solid circles) were taken from the calculations. %In Fig. \ref{fig:naqmom}, the solid circles are the present data, for which signs are taken from the calculations, the open circles are the known experimental data and the theoretical calculations are connected by the solid line.  
Calculations with the USD \cite{wildenthal83} or the USDB \cite{brown06} interactions gave results within the expected theoretical uncertainty and therefore only the results of the calculations performed with USDA interaction are discussed here.  

The general trend of $Q$ of $^{20-27}$Na isotopes is well reproduced by the shell-model calculations.  In particular, the $Q$ of the neutron-deficient nucleus $^{21}$Na and even of the proton drip line nucleus $^{20}$Na are approximately accounted for by the USDA interaction \cite{brown06}, which is determined by fitting energy levels of only neutron-rich nuclei in the $sd$ shell.  The present $Q(^{21}$Na) reconciles the discrepancy between the previous $Q(^{21}$Na) and the theoretical calculation, and completes the account of the systematic behavior of $Q$ of Na isotopes across the neutron 1$d_{5/2}$ and 2$s_{1/2}$ shells except $Q(^{24}$Na), which is not measured yet.    

No variation of effective charges were required to realize agreement between the shell-model calculations in the $sd$-shell model space and experiment.  This was most surprising for the case of $^{20}$Na, which lies at the proton drip line.  Nuclei adjacent to both the proton and neutron drip lines have been shown to require significantly different $e_p$ and $e_n$ values to reproduce experimental $Q$.  The neutron-rich B isotopes, as discussed in section \ref{section:1}, show a reduction of the neutron effective charge \cite{ogawa04} %because the loosely-bound valence neutrons, far removed from the core, have less probability to excite the core than well bounded neutrons.  
and neutron-deficient nucleus $^{37}$K requires an increased neutron effective charge and a decreased proton effective charge \cite{minamisono08b}.% due to a substantial coupling to the isovector giant resonance beyond that indicated by the typical effective charges. 

The isovector part, $e_{\rm pol}^{(1)}$, of effective charges was varied to investigate its contribution to the present $Q$, keeping the isoscalar part, $e_{\rm pol}^{(0)}$, the same as the one determined from the $E$2 matrix elements between low-lying states \cite{brown88}, which are sensitive to the $e_{\rm pol}^{(0)}$ but less sensitive to the $e_{\rm pol}^{(1)}$.  Here $e_{\rm pol}^{(0)}$ and $e_{\rm pol}^{(1)}$ are defined as $e_p = 1e+e_{\rm pol}^{(0)}-e_{\rm pol}^{(1)}$ and $e_n = e_{\rm pol}^{(0)}+e_{\rm pol}^{(1)}$.  In Fig. \ref{fig:naqmom}, the dashed line is calculated with $e_p$ = 1.4$e$, $e_n$ = 0.4$e$ and the dotted line with $e_p$ = 1.1$e$, $e_n$ = 0.7$e$.  Only a slight variation can be seen for $Q(^{20}$Na).  The $Q$ of Na isotopes are not sensitive to the variation of the $e_{\rm pol}^{(1)}$ within the experimental errors but mainly determined by the $e_{\rm pol}^{(0)}$. The result indicates that the collective behavior of protons and neutrons dominate the $Q$ of Na isotopes.  This is also implied by the similar value of proton- and neutron-bare quadrupole moments of each Na isotopes listed in Table \ref{table:theory}.   

Results of shell-model calculations from $A$ = 20 to 25 with valence nucleons restricted in the 1$d_{5/2}$ shell are shown in Fig. \ref{fig:naqmom} in the dotted-dashed line.  In the calculation, the USDA interaction and Woods-Saxson single-particle wave functions were used.  The calculated $Q$ are systematically smaller than experimental values and indicate an importance to include the 2$s_{1/2}$ and 1$d_{3/2}$ shells for configuration mixing within the $sd$-shell model space.  By including the 2$s_{1/2}$ and 1$d_{3/2}$ shells in the calculation, each component of the wave function contributes coherently to $Q$ and adds up to the theoretical $Q$, which is discussed above and shown in the solid line in Fig. \ref{fig:naqmom}.  This is again an indication of the collective nature of $Q$ of Na isotopes ($A$ = 20 $\sim$ 24).  The small $Q$ of $^{25, 26, 27}$Na are well explained in the framework of the single-particle picture as follows.  The 3 valence protons, occupying half the 1$d_{5/2}$ shell, yield zero $Q$ \cite{castel90}.  The 6 neutrons in $^{25}$Na, that fully occupies the 1$d_{5/2}$ shell, couples to 0$^+$ and the 7th and 8th valence neutrons in $^{26, 27}$Na occupy the 2$s_{1/2}$ shell.  

A ratio of $Q$ between $^{20}$Na and $^{21}$Na can be precisely determined from the present $eqQ/h$ [Eqs. (\ref{eq:couplingconst20}) and (\ref{eq:couplingconst21})] without any ambiguities, possibly caused by the extraction of $eqQ(^{27}{\rm Na})/h$ [Eq. (\ref{eq:couplingconst27})] and $Q(^{27}$Na) \cite{keim00}:
\begin{equation}
\left[ \frac{eqQ(^{20}{\rm Na})}{h} \bigg/ \frac{eqQ(^{21}{\rm Na})}{h} \right]_{\rm ZnO} = \frac{Q(^{20}{\rm Na})}{Q(^{21}{\rm Na})} = 0.73 \pm 0.02.
\end{equation} 
The ratio clearly indicates that the $Q(^{21}$Na) is larger than $Q(^{20}$Na).  The ratio is consistent with the trend of the $Q$ predicted by shell-model as seen in Fig. \ref{fig:naqmom}, whereas the previous experimental values as well as the shell-model calculations with valence nucleons restricted in the 1$d_{5/2}$ shell [the dotted-dashed line in Fig. \ref{fig:naqmom}] show the opposite trend.  The present ratio of $eqQ/h$ reconciles the disagreement, at least, in the trend of previous experimental and theoretical $Q$, and are not explained by the single-particle picture.

\section{SUMMARY}
Nuclear polarized beams produced by the optical pumping technique were used at ISAC/TRIUMF to determine the quadrupole moments of the proton drip line nucleus $^{20}$Na and the neutron-deficient nucleus $^{21}$Na.  The electric-quadrupole coupling constants of $^{20}$Na and $^{21}$Na, implanted in a hexagonal ZnO single crystal, were measured to be $|eqQ/h|$ = 690 $\pm$ 12 kHz and $|eqQ/h|$ = 939 $\pm$ 14 kHz, respectively, using the multi-frequency $\beta$-ray detecting nuclear magnetic resonance technique under presence of an electric-quadrupole interaction.  An electric-quadrupole coupling constant of $^{27}$Na in ZnO was also measured as $|eqQ/h|$ = 48.4 $\pm$ 3.8 kHz.  The quadrupole moments of $^{20}$Na and $^{21}$Na were deduced as $|Q(^{20}{\rm Na})|$ = 10.3 $\pm$ 0.8 $e$ fm$^2$ and $|Q(^{21}{\rm Na})|$ = 14.0 $\pm$ 1.1 $e$ fm$^2$, respectively, using the $eqQ(^{27}{\rm Na})/h$ and known quadrupole moment of $^{27}$Na as references.  The present $Q(^{20}$Na) is consistent and as precise as the previous report.  The present $Q(^{21}$Na) is not consistent though more precise than the previous value (6 $\pm$ 3 $e$ fm$^2$) obtained from the hyperfine coupling constant measured by laser spectroscopy.  Both of the preset $Q$ are well reproduced by shell-model calculations in the full $sd$-shell model space with USDA interaction and using typical effective charges in the $sd$ shell, $e_p = 1.3e$ and $e_n = 0.5e$.  %No variation of the effective charges from the typical values are required even for these neutron deficient $^{20}$Na and $^{21}$Na isotopes.  
The present $Q(^{21}$Na) reconciles the deviation between the previous $Q(^{21}$Na) and the theoretical calculation, and gives full account of systematic behavior of $Q$ of Na isotopes across the neutron 1$d_{5/2}$ and 2$s_{1/2}$ shells except $Q(^{24}$Na), which is not measured yet.  The shell model shows that the collective behavior of protons and neutrons dominate present $Q(^{20}$Na) and $Q(^{21}$Na). %The shell model shows that contributions to the $Q$ from different component of the wave function within the $sd$-shell model space coherently adds up to the theoretical $Q$, whereas calculations with valence nucleons restricted in the 1$d_{5/2}$ shell do not reproduce experimental values.  The collective behavior of protons and neutrons dominate present $Q(^{20}$Na) and $Q(^{21}$Na).  

\section*{ACKNOWLEDGMENT}
The authors would like to thank P. F. Mantica and B. A. Brown at NSCL/MSU for valuable discussions and are grateful to the TRIUMF staff.  The study was supported in part by the 21th century COE program 'Towards a new basic science: depth and synthesis'.  One of the authors (K. M. at NSCL/MSU) would like to thank the support from the National Science Foundation, Grant PHY06-06007.

% \appendix

% \section{}
% \label{}

\newpage

\newpage

\newpage

\newpage

\newpage

\newpage

\newpage

\begin{table}
\begin{center}
\caption{Experimental conditions in the $eqQ/h$ measurements of $^{20, 21, 27}$Na.  $P_0$ is the initial polarization, $T_1$ is the relaxation time of polarization, $E_0$ is the maximum $\beta$-decay energy of the largest branch, $B_r$, and $A_s$ is the $\beta$-decay asymmetry parameter integrated over measured branches.}
\begin{tabular}{lccc}
\hline%
& $^{20}$Na & $^{21}$Na & $^{27}$Na\\
\hline%
$I^{\pi}$ & 2$^{+}$ & 3/2$^{+}$ & 5/2$^{+}$\\
$T_{1/2}$ (ms) & 447.9 & 22490 & 301\\
$P_0$ (\%)  in ZnO & 37 $\pm$ 1 & 29.7 $\pm$ 0.4 & 39.9 $\pm$ 2.5\\
$T_1$ (s) in ZnO & 9.0 $\pm$ 0.5 & 9.63 $\pm$ 0.09 & 9.5 $\pm$ 2.5\\
%decay mode & $\beta^+$ & $\beta^+$ & $\beta^-$\\
$E_0$ (MeV) & 11.23 & 2.53 & 8.08\\
$B_r$ (\%) & 79.3 & 94.97 & 84.3\\
$A_s$ & +0.33 & +0.81 & -0.88\\
FM (kHz) & $\pm$ 20 & $\pm$ 50 & $\pm$ 20\\
RF time (ms) & 10 & 20 & 10\\
$H_1$ (mT) & 0.8 & 0.65 & 0.8\\
\# of transition freq. & 4 & 3 & 5\\
\# of RF for AFP & 10 & 6 & 15\\
\hline%
\end{tabular}%
\label{table:expcondition}
\end{center}
\end{table}

%\newpage

\begin{table}
\begin{center}
\caption{Present $eqQ/h$ and resulting $Q$ of $^{20, 21, 27}$Na.  Previously known $Q$ and theoretical predictions are also summarized.  $e_p$ = 1.3$e$ and $e_n$ = 0.5$e$ are used for calculations.}
\begin{tabular}{lccc}
\hline
 & $^{20}$Na & $^{21}$Na & $^{27}$Na \\ 
\hline
$|eqQ/h| _{\rm ZnO}$ (kHz) & $689.6 \pm 2.0 \pm 11.8$ & $939.3 \pm 1.9 \pm 13.7$ & 48.4 $\pm$ 2.4 $\pm$ 3.0\\
$|Q/Q(^{27}{\rm Na})|$ & $14.3 \pm 1.2$ & 19.4 $\pm$ 1.6 & 1 \\
$|Q|$ ($e$ fm$^2$) & 10.3 $\pm$ 0.8  & 14.0 $\pm$ 1.1 & 0.72 $\pm$ 0.03 \cite{keim00}\\
$Q_{\rm literat.}$ ($e$ fm$^2$) & 9 $\pm 1$ \cite{keim00b} & $+5.8 \pm 3.3$ \cite{touchard82,wo93,jonsson96} & $-$0.72 $\pm$ 0.03 \cite{keim00}\\
$Q_{\rm theory}$ ($e$ fm$^2$) \cite{oxbash}& +9.9 & +11.6 & $-$1.1 \\
\hline \\
\end{tabular}%
%\footnotetext[1]{present work.}
%\footnotetext[2]{from Ref. \cite{keim00b}.}
%\footnotetext[3]{evaluated using Ref. \cite{touchard82,wo93,jonsson96}.}
%\footnotetext[4]{from Ref. \cite{keim00}.}
%\footnotetext[5]{calculated by OXBASH \cite{oxbash}.}
\label{table:expresult}
\end{center}
\end{table}

\newpage

\begin{table}
\begin{center}
\caption{Shell-model calculations of $Q$ of Na isotopes.  $Q_{\rm theory}$ were obtained using $e_p$ = 1.3$e$ and $e_n$ = 0.5$e$.  Signs of $Q_{\rm exp.}$ are given when the sign can be determined by the experiment.}
\begin{tabular}{lcccc}
\hline
 & $Q^p_0$ & $Q^n_0$ & $Q_{\rm theory}$ ($e$ fm$^2$) & $Q_{\rm exp.}$ ($e$ fm$^2$)\\
\hline
$^{20}$Na & 6.4 & 3.1 & 9.9 & 10.3 $\pm$ 0.8\\
 & & & & 9 $\pm$ 1 \cite{keim00b}\\
$^{21}$Na & 6.7 & 5.8 & 11.6 & 14.0 $\pm$ 1.1\\
 & & & & 5.8 $\pm$ 3.3 \cite{touchard82,wo93,jonsson96}\\
$^{22}$Na & 12.9 & 12.9 & 23.2 & +18.5 $\pm$ 1.1 \cite{gangrsky98}\\
$^{23}$Na & 5.9 & 6.5 & 10.9 & +10.6 $\pm$ 0.1 \cite{stone05}\\
$^{24}$Na & 14.4 & 12.3 & 24.9 & -\\
$^{25}$Na & $-$0.4 & 1.9 & 0.4 & $-$6.4 $\pm$ 4.4 \cite{touchard82,wo93,jonsson96}\\
 & & & & 0.10 $\pm$ 0.04 \cite{matsuta05}\\
$^{26}$Na & $-$0.5 & 0.4 & $-$0.4 & $-$5.6 $\pm$ 4.7 \cite{touchard82,wo93,jonsson96}\\
 & & & & $-$0.53 $\pm$ 0.02 \cite{keim00}\\
$^{27}$Na & $-$0.2 & $-$1.7 & $-$1.1 & $-$2.9 $\pm$ 4.5 \cite{touchard82,wo93,jonsson96}\\
 & & & & $-$0.72 $\pm$ 0.03 \cite{keim00}\\
\hline
\end{tabular}
\label{table:theory}
\end{center}
\end{table}

\end{document}